# Fixed Pattern Noise Reduction for Infrared Images Based on Cascade Residual Attention CNN

Juntao Guan[a], Rui Lai[a,*], Ai Xiong[b], Zesheng Liu[a], Lin Gu[c]

[a]*Department of Microelectronics, Xidian University, Xi'an 710071, China*
[b]*School of Control Engineering, Chengdu University of Information Technology, Chengdu 610103, China*
[c]*National Institute of Informatics (NII), Tokyo 101-8430, Japan*

**Abstract**

Existing fixed pattern noise reduction (FPNR) methods are easily affected by the motion state of the scene and working condition of the image sensor, which leads to over smooth effects, ghosting artifacts as well as slow convergence rate. To address these issues, we design an innovative cascade convolution neural network (CNN) model with residual skip connections to realize single frame blind FPNR operation without any parameter tuning. Moreover, a coarse-fine convolution (CF-Conv) unit is introduced to extract complementary features in various scales and fuse them to pick more spatial information. Inspired by the success of the visual attention mechanism, we further propose a particular spatial-channel noise attention unit (SCNAU) to separate the scene details from fixed pattern noise more thoroughly and recover the real scene more accurately. Experimental results on test data demonstrate that the proposed cascade CNN-FPNR method outperforms the existing FPNR methods in both of visual effect and quantitative assessment.

## 1. Introduction

Infrared focal plane arrays (IRFPAs), being the visual sensors of staring infrared imaging systems, have been widely used in various military and civilian applications [1-3]. However, IRFPAs are affected by the fixed pattern noise (FPN), which is mainly caused by the spatial non-uniform response of individual detectors in the sensor [6-7]. More seriously, spatial FPN generally drifts with time, which makes the problem be more challenging [8-11]. As a result, the FPN causes a significant decline in imaging quality and decreases the precision for object detection and recognition. To meet this challenge, the cost-effective fixed pattern noise reduction (FPNR) techniques based on signal processing are continually investigated and applied in nearly all the infrared imaging systems.

* Corresponding author.
*E-mail address:* rlai@mail.xidian.edu.cn.(Rui Lai)

Existing FPNR algorithms are mainly divided into two primary categories: reference-based FPNR (RB-FPNR) and scene-based FPNR (SB-FPNR) [12-14]. The RB-FPNR methods remove the FPN according to fixed calibration parameters calculated from the response of blackbody radiation at different temperatures [15]. Unfortunately, such a calibration requires the camera to halt the normal operation and update the calibration parameters due to the inherent temporal drift of detector characteristics [16]. Given this fact, most of the recent researches have focused on developing SB-FPNR methods, such as neural networks (NN) [17], temporal high-pass filter (THPF) [18, 19] and constant-statistics (CS) method [20-21]. As for SB-FPNR algorithms, the calibration parameters are iteratively updated by utilizing the information extracted from inter-frame motion, therefore, ghosting artifacts and over smooth effects resulted from the sudden deceleration of scene motion often seriously degrade the noise reduction performance, moreover, the relatively slow convergence process occurred in scene switching is unacceptable for most of the practical applications.

In recent years, convolution neural network (CNN) [22] models were explored deeply and applied in various image processing tasks [23], such as image super resolution [24, 25], image denoising [26], and sketch synthesis [27-29]. To the best of our knowledge, CNN based FPNR methods still have not been extensively investigated in the literature. Current deep learning based methods only focus on the stripe noise removal problem [30, 31]. Motivate by the existing researches on CNN, we present a cascade CNN model and develop a corresponding FPNR method, which can continuously realize the single frame blind FPNR operation without any parameter tuning and suppress ghosting artifacts as well as eliminate the over smooth effect simultaneously. The proposed innovative model is trained in an end-to-end fashion and directly learns the calibration parameters for image restoration from the noise corrupted images.

The major contributions of our work are briefly summarized as follows:

1. Propose an innovative cascade CNN model to directly learn calibration parameters from training data rather than relying on predefined priors or filters, which yields higher noise suppression accuracy without complicated parameter tuning as well as avoids ghosting artifacts and over smooth effects.

2. Design a coarse-fine convolution (CF-Conv) unit to extract various coarse and fine grained features efficiently. This technique integrates more spatial information from a various sized receptive field and utilizes the complementarity of different features to improve the accuracy of FPNR method.

3. Construct a spatial-channel noise attention unit (SCNAU) in the newly presented CNN model according to visual attention mechanism. The SCNAU separates the scene detail related features from the FPN related features more thoroughly to recover the real scene more accurately.

The remainder of this paper is organized as follows. Section 2 gives an overview of the related works. The principle of the proposed CNN-FPNR method is detailed in Section 3. Thereafter, the experimental results and analysis are presented in Section 4. Finally, the conclusions of this research are stated in Section 5.

## 2. Related Works

### 2.1. Scheme of FPNR

The response of each detector in IRFPA can be represented by the following linear model

$$y_{ij} = g_{ij} \cdot x_{ij} + o_{ij} \tag{1}$$

where $x_{ij}$ and $y_{ij}$ are the actual radiation and observed response of the $(i,j)$ detector, $g_{ij}$ and $o_{ij}$ denote the gain noise and offset noise, respectively [32].

The goal of the FPNR is to remove the FPN from the observed response and calculate the actual radiation with the estimated calibration parameters. According to the linear response model defined in Equation (1), the FPNR model can be represented as

$$\hat{X}_{ij} = \hat{W}_{ij}^T \cdot Y_{ij} \tag{2}$$

where $\hat{X}_{ij}$ indicates the estimated radiation, $Y_{ij} = (y_{ij}, 1)^T$ represents the observation, $\hat{W}_{ij} = (\hat{G}_{ij}, \hat{O}_{ij})^T$ stands for the calibration parameter matrix.

As for RB-FPNR methods, the black body was employed as the reference source to provide the practical radiance in different temperature, and then the response of IRFPA are collected and applied to calculate the calibration parameters, for the widely used two-point calibration, $\hat{W}_{ij}$ can be solved by the linear equation

$$\begin{cases} \overline{Y}^L = \hat{W}_{ij}^T \cdot Y_{ij}^L = \hat{G}_{ij} \cdot y_{ij}^L + \hat{O}_{ij} \\ \overline{Y}^H = \hat{W}_{ij}^T \cdot Y_{ij}^H = \hat{G}_{ij} \cdot y_{ij}^H + \hat{O}_{ij} \end{cases} \tag{3}$$

and the mean response of low and high temperature radiation are given by

$$\overline{Y}^L = \frac{1}{M \cdot N} \sum_{i=1}^{M} \sum_{j=1}^{N} Y_{ij}^L \tag{4}$$

$$\bar{Y}^H = \frac{1}{M \cdot N} \sum_{i=1}^{M} \sum_{j=1}^{N} Y_{ij}^H \qquad (5)$$

where $Y_{ij}^L$ and $Y_{ij}^H$ represent the response of low and high temperature scene, respectively.

For SB-FPNR methods, the $\hat{W}_{ij}$ is generally estimated by minimizing the following objective function

$$J_{ij} = \alpha \cdot \left\| \hat{X}_{ij} - T_{ij} \right\|_2^2 + \lambda \cdot \left\| \Phi(\hat{X}_{ij}) \right\|_p^p \qquad (6)$$

where $\alpha$ and $\lambda$ are the scalar for properly weighting the fidelity term against the regularization term, the target image $T_{ij}$ is acquired by the local spatial filtering approach, and $p$ indicates the norm of the penalty function.

For the NN-FPNR [17] and FA-FPNR [33] methods, the objective function is simplified by assigning $\lambda = 0$. On the contrary, the fidelity term is removed by setting $\alpha = 0$ and the penalty term is formulated as $\Phi = \nabla$ in the TV-FPNR [34] method.

To minimize the objective function $J_{ij}$, the steepest descent algorithm is used to iteratively search the optimum solution of calibration parameters formulated by

$$\hat{W}_{ij}^{n+1} = \hat{W}_{ij}^n + \mu_{ij} \cdot D\left(J_{ij}^n\right) \qquad (7)$$

where $\mu_{ij}$ represents the learning rate, $D(\cdot)$ denotes the partial derivative operation.

Distinguishingly, the proposed CNN-FPNR method employs CNN to directly estimate the calibration parameter $\hat{W}_{ij}$ from single observation $Y_{ij}$ without any pre-calibration or iteratively optimization.

*2.2. Visual Attention Mechanism*

One important property of human visual perception is that one tends to avoid processing a whole image at once and focuses attention selectively on parts of the visual space to acquire information when and where it is needed, and then combines information from different fixations over time to build up an internal decision making for the image [35]. Enlightened by the human perception process, several tentative efforts have applied attention mechanism to structural prediction tasks such as image or video classification, image segmentation [36] and image caption [37]. However, recent works use hard pooling to select the most probably attentive region in the spatial domain, which is liable to lose detail information and decrease attention accuracy. More seriously, this attention strategy is only applied in the last convolution layer and limits the receptive field, which inevitably weakens the capability of spatial attention [38-39]. Besides the spatial attention, Chen et al. proposed a semantic attention mechanism to select the relevant features based on context semantics in image caption tasks [40]. As for FPNR tasks, the feature reflecting the characteristics of noise and image details are mixed together, which seriously disturbs the estimation of calibration parameters. Therefore, it is necessary to explicitly model the interdependence between feature channels for selecting the task related features, which will be of benefit to improving the performance of the CNN model.

In view of this, a novel attention mechanism will be proposed to pay more attention to the noise related pixels and features in both of spatial and channel domain to improve the calibration accuracy.

*2.3. Dilation and Sub-pixel Convolution*

Dilation convolution generally employed to enlarge the receptive filed and extract the coarse-grained information without additional computational complexity. Lai [41] showed that dilated convolutions were helpful for pixel-level segmentation tasks. DCSR [42] introduced the dilation convolution in image super resolution tasks and achieved effective results. However, dilation operator often produces blind spots in the receptive filed, which is caused by adopting the same filter with various dilation factors at different ranges.

As for fine-grained contexture extraction, the sub-pixel convolution fuses multi-channel features to extract the sub-pixel information and reposition the features in pixel level [43]. The sub-pixel convolution has been investigated and utilized to combat the checkboard artifacts appeared in flow estimation and semantic segmentation [44].

In view of this, we propose a mixed convolutional layer combining dilated convolutions, sub-pixel convolutions and standard convolutions to extract the multi-grained features for improving the calibration accuracy.

**3. Proposed Method**

In order to realize the scene adaptive single frame FPNR and eliminate the ghosting artifacts as well as over smooth effects, we proposed a cascade residual CNN-FPNR method. As shown in Figure 1, the two-stage cascade residual CNN model is constructed to implement the end-to-end FPNR for infrared image. Briefly speaking, the presented CNN model sequentially estimates the gain and offset calibration parameters from observation $Y_{ij}$ by using deep learning strategy, and then uses the estimated calibration parameters to estimate the actual response $\hat{X}_{ij}$ according to the calibration model mathematically expressed by Equation (2).

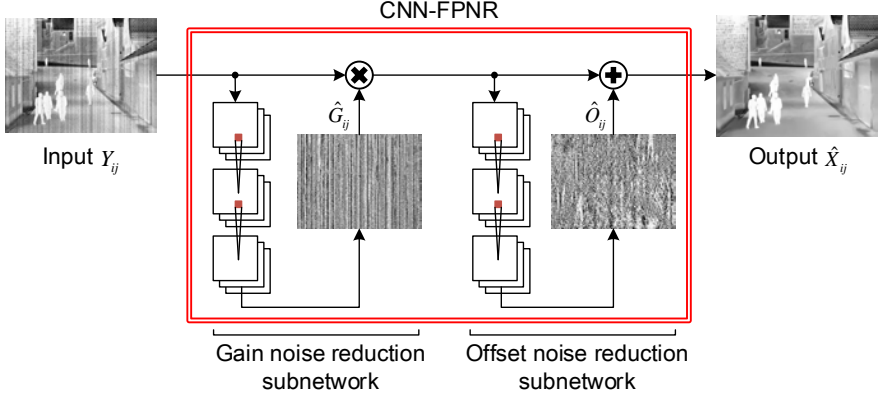

**Figure 1.** Scheme of the proposed cascade residual CNN-FPNR method

*3.1. Architecture of the Proposed CNN Model*

As mentioned in Section 2.1, the FPN consists of two different types of noise. Removing the mixed FPN through single CNN model would inevitably lead to detail loss in the calibrated results. In view of this, two noise reduction subnetworks are cascaded to respectively remove the gain noise and offset noise, the whole network architecture is shown in Figure 2. Moreover, since the input image and output image are highly correlated, it is sufficient to directly estimate the calibration parameters from the input image using residual learning strategy. In each noise reduction subnetwork, five feature extraction blocks (FEBs) consists of CF-Conv unit and SCNAU are stacked to refine the noise related features, while the last convolution layer consists of a single filter with kernel size 1×1 is introduced to fuse channel wise information and output the estimated calibration parameters.

In gain noise calibration subnetwork, the residual layer with skip connection multiply the input image by the estimated gain calibration parameter $\hat{G}_{ij}$ to remove the fixed stripe gain noise, which can be formulated as

$$C_G = \hat{G}_{ij} \odot Y_{ij} \tag{8}$$

where $C_G$ indicates the gain noise calibrated image, $\odot$ denotes the dot product operator.

For the offset noise calibration, the calibration parameter $\hat{O}_{ij}$ is added to the gain noise calibrated image $C_G$ to further suppress the additive offset noise, which can be represented as

$$\hat{X}_{ij} = C_G + \hat{O}_{ij} \tag{9}$$

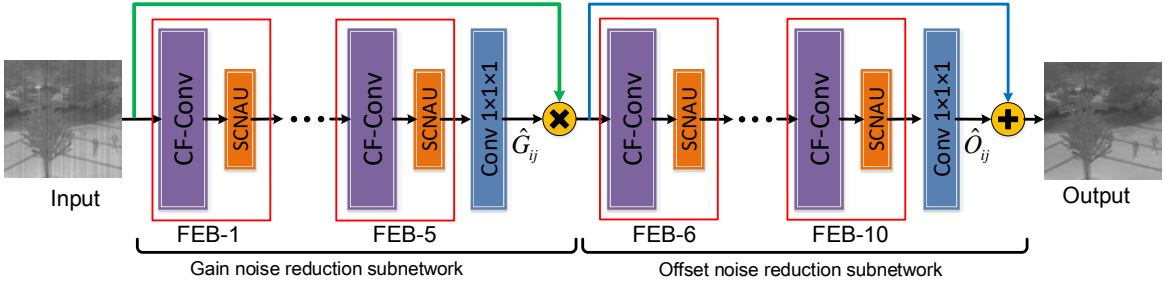

**Figure 2.** The network architecture of the proposed CNN-FPNR method

*3.2. Coarse-Fine Convolution Unit*

As for convolutional network based pixel-level prediction problems, the restoration performance is highly influenced by the rich contextual information extracted by convolutional kernels. However, it is worthless to directly enlarge the receptive field for capturing larger scale contextual information, which will cause additional computational complexity and memory consumption [43]. In this work, we propose a CF-Conv unit to pick and fuse rich multi-grained spatial wise features comprehensively. As can be seen in Figure 3, dilation convolution (Dia-Conv) filters and sub-pixel convolution (SP-Conv) filters are employed to extract the coarse-grained and fine-grained features, respectively. The blind spots caused by dilation convolution are filled by standard convolution (Std-Conv). All the outputs of the convolution layer are activated with ReLU function and concatenated as a single output vector. Then, a bottleneck convolution layer Std-Conv-2 is introduced to perform feature reduction and fuse cross-grained information [45, 46]. The design follows the practical intuition that visual information should be processed at various granularities and then aggregated. This technique will be meticulously discussed in Section 4.2.

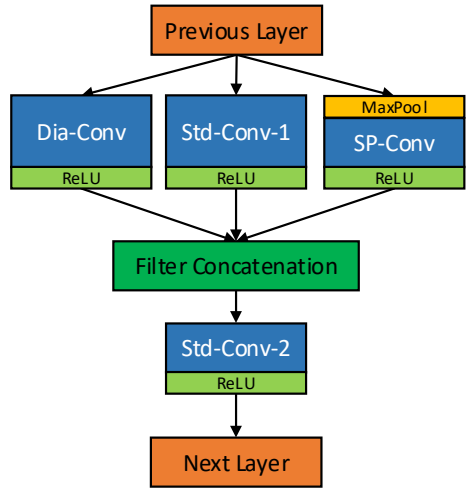

**Figure 3.** The detail architecture of CF-Conv unit

The detail configuration of the CF-Conv unit is given in Table 1. According to previous experiments, the 3×3 sized filter shows excellent feature extraction ability in image super resolution and denoising tasks [45]. Therefore, the filter size of CF-Conv unit is restricted to 3×3. Moreover, zero-padding was applied to make sure the output of each layer shares the same dimension with the input feature.

**Table 1.** The configuration of proposed MsConv unit

| Layer | Layer Type | Parameters | Filter Number | Activation Function | Stride |
|---|---|---|---|---|---|
| MaxPool | Max Pooling | Pooling Size = 2 | -- | -- | 2, 2 |
| Dia-Conv | Dilation Convolution | Dilation Rate = 2 | 32 | ReLU | 1, 1 |
| Std-Conv-1 | Standard Convolution | -- | 64 | ReLU | 1, 1 |
| SP-Conv | Sub-pixel Convolution | Sub-pixel Scale = 4 | 32 | ReLU | 1, 1 |
| Std-Conv-2 | Standard Convolution | -- | 64 | ReLU | 1, 1 |

*3.3. Spatial-Channel Noise Attention Unit*

In FPNR tasks, the feature map directly extracted by CNN is the aggregate of image detail and fixed pattern noise related features, which will cause detail loss in the calibrated results. In view of this, we will propose a spatial-channel attention mechanism to separate the mixed features to improve correction accuracy. Recent works have demonstrated the performance of CNN can be improved by explicitly embedding visual attention mechanisms, which helps to capture spatial correlations without requiring additional supervision [36]. The attention mechanisms can be categorized as spatial domains and channel domains. The spatial attention mechanism attempts to pay more attention to the noise-related regions instead of considering each feature map equally. In other words, the spatial attention serves as a pattern selector and depresses the clutter background and unconcerned features in the spatial domain. Unlike spatial attention, the channel attention mechanism performs as a feature selector and refines the features by assigning dynamic weight for different channels. Specially speaking, channel attention mechanism can be applied to enhance the noise related features and suppress the detail related features for estimating the calibration parameters more precisely.

Inspired by the recent visual attention mechanism, we present noise feature refining SCNAU in this paper. As shown in Figure 4, each SCNAU consists of three branches called spatial attention branch, trunk branch and channel attention branch, respectively. Specifically, spatial attention branch is constructed with stacked convolutional layers, and the trunk branch is carried out by identity mapping. In addition, the channel attention branch consists of a global average pooling layer followed by a set of full connect operation. The detail configuration of the SCNAU is listed in Table 2. Worthy of note is the last activation layer in each attention branch is activated by the sigmoid function, which normalizes the output to range [0, 1]. The mechanism of SCNAU can be mathematically represented as

$$x_{sc} = f_s(x) \odot f_c(x) \otimes x \tag{10}$$

where $f_s$ stands for the spatial attention mask, the $f_c$ means the channel attention mask, $f_s(x) \odot f_c(x)$ represents the spatial-channel mask responding to the input $x$, the $x_{sc}$ indicates the refined features. The operator $\odot$ and $\otimes$ denotes as the pixel wise multiplication and the channel-wise multiplication, respectively.

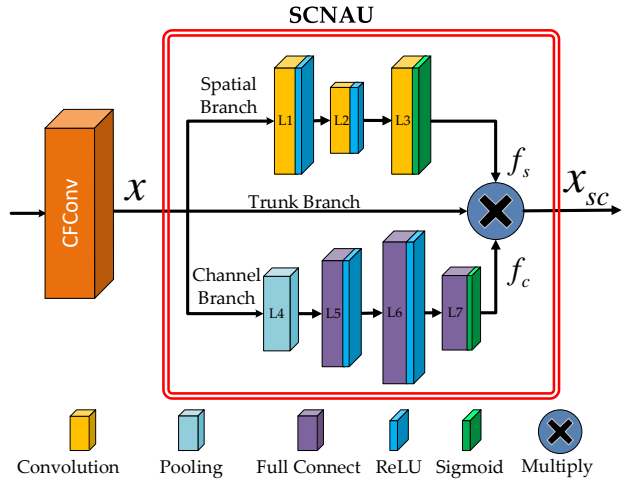

**Figure 4.** The architecture of spatial and channel noise attention unit (SCNAU)

**Table 2.** The detail configuration of spatial channel noise attention unit (SCNAU)

| Layer | Layer Type | Kernel size | Stride | Activate function |
|---|---|---|---|---|
| L1 | Convolution | 64×3×3×64 | 1, 1 | ReLU |
| L2 | Convolution | 64×3×3×32 | 1, 1 | ReLU |
| L3 | Convolution | 32×1×1×64 | 1, 1 | Sigmoid |
| L4 | Global Average Pooling | 2×2 | 2, 2 | -- |
| L5 | Full Connected | 64×256 | -- | ReLU |
| L6 | Full Connected | 256×521 | -- | ReLU |
| L7 | Full Connected | 521×64 | -- | Sigmoid |

Figure 5 shows how the SCNAU implement the separation of scene detail and noise related features with different noise attention masks. It is clear that the detail related features such as the edge of eaves and human body contour are mixed in the original noise feature map. If the noise feature map doped scene detail features is employed to produce calibration parameters, the CNN-FPNR method without SCNAU will discard useful scene details in the calibration results. In view of this, we utilize the spatial-channel noise masks to diminish the scene detail related features from original noise feature map and screen out more pure noise features. As can be seen from Figure 5, the masked features shown in the last column really involves far less scene details when compared with the original noise feature map.

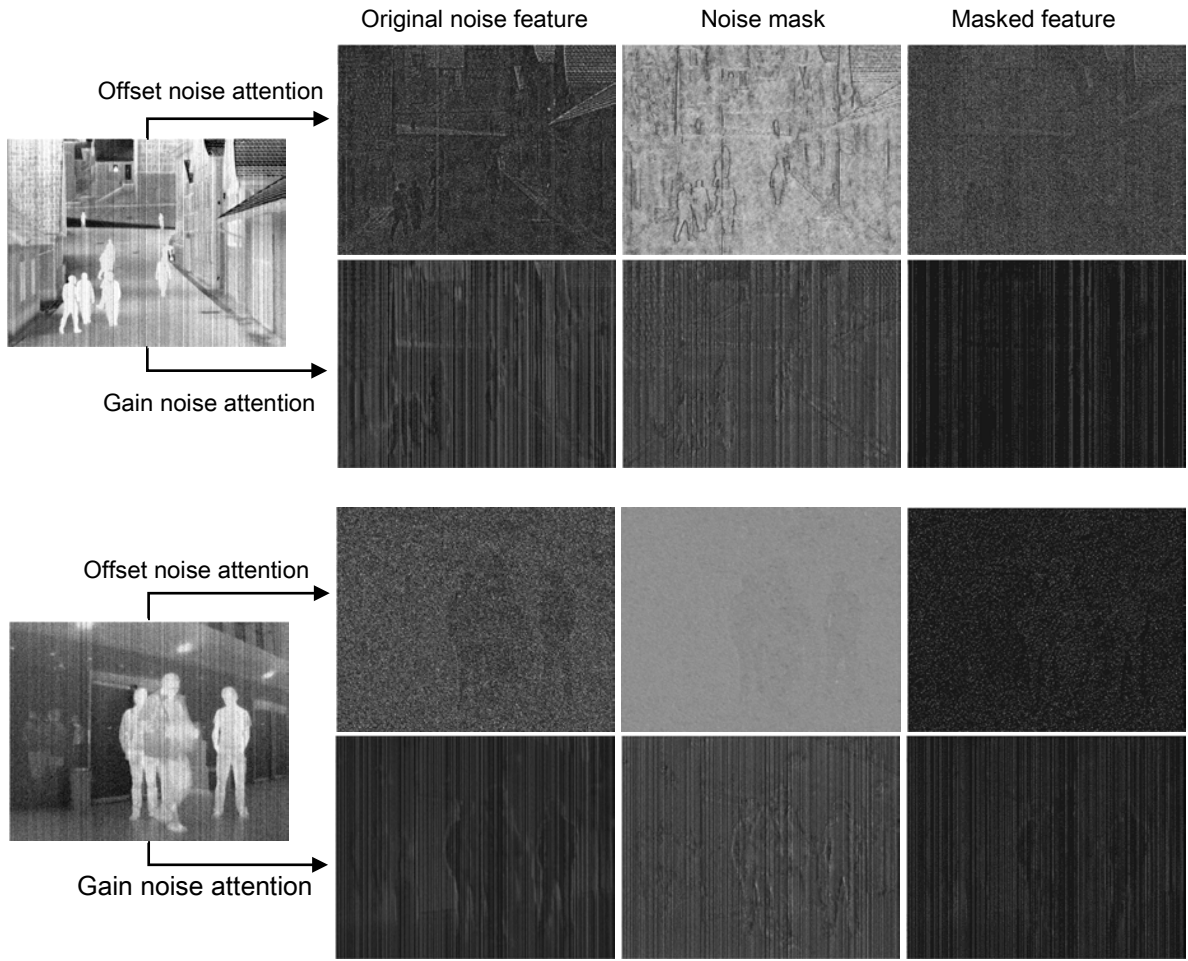

**Figure 5.** Illustration of the noise attention mechanism

## 4. Experimental Results and Analysis

In this section, we will describe the training strategy and demonstrate the effectiveness of the proposed CF-Conv and SCNAU for noise reduction, the related experiments are applied on an independent testing data set Set12 which was collected in [25]. Subsequently, we will verify the effectiveness of our proposed CNN-FPNR method on both of artificially corrupted simulation data and real noisy data with the traditional scene based NN-FPNR [17], FA-FPNR [33], TV-FPNR [34] method and deep learning based DLSNUC [30] and ICSRN [32] method. It is worthy to note that the DLSNUC and ICSRN focus to suppress the stripe nonuniformity. Therefore, we retrain the DLSNUC and ICSRN to handle the mixed noise scenario. As for the simulation data, 4000 frames 471×358 sized and 1700 frames 312×384 sized infrared images artificially corrupted with FPN and respectively called sequence 1 and sequence 2 are utilized to evaluate the calibration performance. Moreover, 1970 frames of 384×288 sized and 200 frames of 201×181 sized real noisy infrared images are collected and respectively named sequence 3 and sequence 4, which are then used to assess the practical performance. For all the competition methods, parameters are properly tuned to pursuit the best performance.

*4.1. Training*

As stated in the literatures, deep learning generally benefits from the large training dataset. To generate a complete training dataset, we select 300,000 frames of 40×40 sized visible images from BSD500 datasets and augmented with horizontal flip and 3 separate angles (90, 180 and 270°) of rotations, and then corrupt them using gain noise with mean 1 and standard deviation from 0.05 to 0.15 as well as offset noise with mean 0 and standard deviation from 5 to 25 according to Equation 2.

For the test dataset, Set12 dataset was used for thorough evaluation, it is worthy of note that the test dataset is widely used to be benchmark in most image denoising and restoration works and not included in the training dataset. In order to investigate the noise reduction performance under different noise strength, gain noise with mean 1 and standard deviation of 0.04, 0.08, 0.12 and offset noise with mean 0 and standard deviation of 5, 10, 15 are respectively applied on the test dataset.

Followed that, the proposed CNN model for blind FPNR task is trained over 50 epochs using the adaptive moment estimation (ADAM) optimization method [47] with mini batch 128. The learning rate is initially set to 0.001 and then decreased by the factor of 10 every 25 epochs. For weight initialization, we use the 'he_normal' initializer described in [48]. All experiments are carried out in Keras package with the tensorflow backend. The training takes about 10 hours on two NVidia 1080Ti GPUs.

*4.2. Analysis of Network Architecture*

In this section, we perform an ablation study to demonstrate the effectiveness of the proposed CF-Conv unit and SCNAU in the proposed method. Each of the components is added one by one to the network and the results for each configuration are involved in the comparison.

*4.2.1 Effectiveness of the CF-Conv Unit*

To demonstrate the effectiveness of the CF-Conv unit, we arrange special experiments to compare the proposed CF-Conv unit with the traditional single-scale 3×3 (Conv(3)) as well as the multi-scale 3×3+5×5 (Conv(3-5)) and 3×3+5×5+7×7 (Conv(3-5-7)) convolution filters by implementing the same blind FPNR task on the same test dataset.

Figure 6 shows the PSNR curves for each configuration of convolution filters. As can be seen, the traditional multi-scale Conv(3-5) and Conv(3-5-7) achieve higher PSNR than single-scale Conv(3) but suffer from lower coverage speed. In contrast, the proposed CF-Conv unit converges more swiftly and achieves over 1dB promotion in PSNR than any other convolution filters. This phenomenon validates that the dilation, sub-pixel and standard convolutions play a complementary roles in feature extraction.

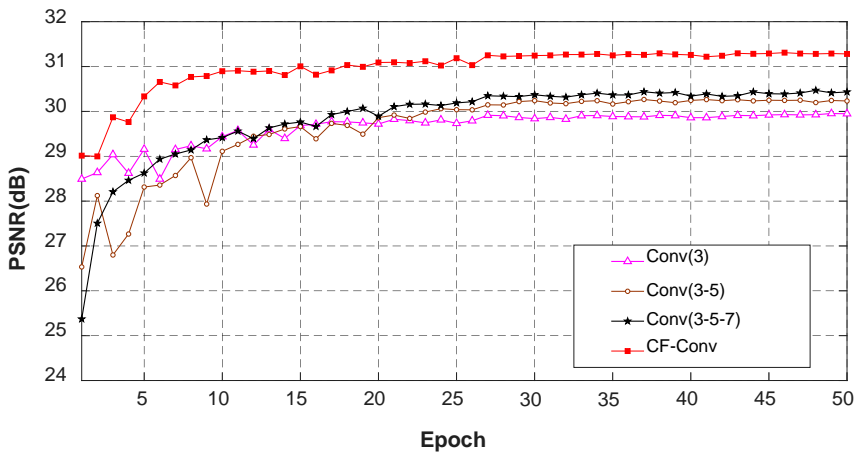

**Figure 6.** PSNR curves of various convolution filters

*4.2.2 Effectiveness of SCNAU*

In order to verify the effectiveness of the SCNAU, we implement the proposed FPNR method on the test dataset using the CNN model with different collocations of attention architecture. The PSNR curves of CNN-FPNR model with Spatial-Channel NAU (SCNAU), Spatial NAU (SNAU) and Channel NAU (CNAU) are shown in Figure 7. It is clear that SNAU and CNAU achieve higher PSNR than the baseline model without NAU, such a result embodies the effectiveness of the attention mechanism. Furthermore, the SCNUA reaches the highest PSNR over other methods. The reason lies in that the SCNAU extracts noise related feature maps from both of spatial and channel perceptive which is beneficial to separate the FPN and the scene detail related features more accurately. The SCNAU makes the proposed CNN-FPNR method maintain a better balance between the FPN suppression and scene detail preservation, which directly leads to a promotion of PSNR over 0.5dB.

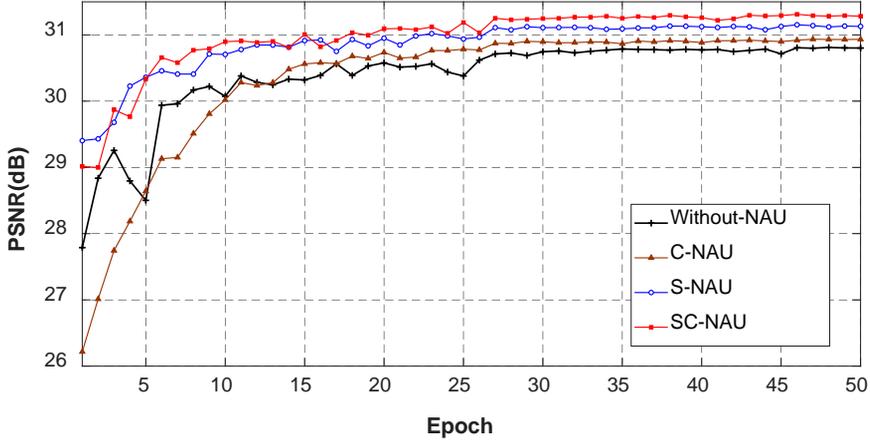

**Figure 7.** PSNR curves of the proposed CNN model with and without NAU

*4.3. Simulation with Artificially Corrupted Image Sequences*

In this section, sequence 1 is artificially corrupted by FPN according to Equation (1), which is then employed to evaluate the calibration accuracy and convergence rate of different FPNR methods. In order to assess the calibration accuracy in various noise strength, the stripe gain noise with mean 1 and standard deviation $\sigma_g$ from 0.08 to 0.12 is firstly applied to the raw sequence 1, and then the offset noise with mean 0 and standard deviation $\sigma_o$ from 5 to 15 generated as realizations of independent and identically distributed (iid) Gaussian random variables is further added to the stripe gain noise corrupted sequence 1.

*4.3.1. Analysis of Correction Accuracy and Convergence Rate*

In the following experiments, PSNR and roughness index [49] are utilized to assess the quality of the calibrated image. The roughness index $\rho$ is defined as

$$\rho = \frac{\|h_1 \otimes \hat{X}\|_1 + \|h_2 \otimes \hat{X}\|_1}{\|\hat{X}\|_1} \tag{11}$$

where $h_1$ is a horizon vector set as $[1,-1]$, and $h_2 = h_1^T$ is a vertical vector. The operator $\|\cdot\|_1$ indicates the L1-norm operator. A lower roughness and higher PSNR means a better calibration result.

The mean values of PSNR and roughness index for each FPNR method are calculated from all the calibrated frames and listed in Table 3. The best results for each noise level are highlighted in bold. As can be seen, both of the deep learning based DLSNUC and ICSRN outperform the traditional NN-FPNR, FA-FPNR and TV-FPNR. Significantly, the proposed CNN-FPNR method obviously achieves the best PSNR and roughness metrics in each noise level. Moreover, with the increase of noise intensity, the performance of existing SB-FPNR methods decreased significantly. In contrast, the proposed CNN-FPNR achieves relatively minor PSNR reduction without any parameter regulation for changing noise level.

**Table 3.** Mean PSNR (dB) /ρ results of various FPNR methods on test sequence 1.

| Noise | | Corrupted Image | Corrected Image | | | | | |
|---|---|---|---|---|---|---|---|---|
| $\sigma_g$ | $\sigma_o$ | | NN-FPNR | FA-FPNR | TV-FPNR | DLSNUC | ICSRN | Proposed |
| 0.08 | 5 | 28.47/0.1792 | 33.32/0.1383 | 34.54/0.1332 | 34.79/0.1349 | 37.65/0.1032 | 37.75/0.1038 | **38.23/0.1030** |
| 0.08 | 10 | 27.21/0.1996 | 32.82/0.1458 | 34.04/0.1390 | 34.61/0.1368 | 37.23/0.1031 | 37.42/0.1035 | **37.94/0.1026** |
| 0.08 | 15 | 26.02/0.2259 | 32.18/0.1551 | 33.37/0.1467 | 34.46/0.1371 | 36.81/0.1025 | 37.10/0.1029 | **37.63/0.1020** |
| 0.10 | 5 | 25.96/0.2058 | 32.13/0.1430 | 33.29/0.1369 | 33.65/0.1206 | 36.72/0.1034 | 36.94/0.1043 | **37.52/0.1029** |
| 0.10 | 10 | 25.84/0.2159 | 31.76/0.1490 | 32.92/0.1418 | 33.16/0.1213 | 36.53/0.1035 | 36.75/0.1040 | **37.35/0.1021** |
| 0.10 | 15 | 24.81/0.2412 | 31.02/0.1584 | 32.08/0.1499 | 32.44/0.1315 | 36.28/0.1028 | 36.53/0.1033 | **37.15/0.1014** |
| 0.12 | 5 | 24.95/0.2224 | 31.10/0.1470 | 32.35/0.1401 | 32.19/0.1353 | 36.14/0.1038 | 36.36/0.1048 | **37.02/0.1025** |
| 0.12 | 10 | 24.54/0.2340 | 30.76/0.1526 | 31.74/0.1455 | 32.16/0.1349 | 35.97/0.1039 | 36.18/0.1044 | **36.87/0.1016** |
| 0.12 | 15 | 23.66/0.2576 | 30.37/0.1623 | 31.33/0.1541 | 31.42/0.1370 | 35.77/0.1031 | 36.01/0.1038 | **36.74/0.1007** |

Figure 8 and Figure 9 respectively present the PSNR and roughness index curves of each FPNR methods upon sequence 1 artificially corrupted by strong FPN with $\sigma_g$=0.15 and $\sigma_o$=15. As shown in Figure 8, the PSNR for traditional NN-FPNR, FA-FPNR and TV-FPNR method have continued to grow for about 1000 frames before convergence. In contrast, the PSNR curves of the deep learning based methods do not require any convergence stage and achieve an amazing promotion up to 5dB over the traditional SB-FPNR methods. Moreover, it is worthy to note that the proposed CNN-FPNR method achieves nearly 1dB promotion over the deep learning based DLSNUC and ICSNR, which is benefit from the much richer details preserved by using the specially designed CNN model based on the coarse-fine grained feature extraction and spatial-channel attention strategy.

Roughness index generally reflects the effect of calibration from the other perspective. As can be seen clearly from the roughness curves plotted in Figure 9, the NN-FPNR, FA-FPNR and TV-FPNR method decrease slowly and their convergence processes last about 1000 frames. Similar to PSNR curves, the proposed CNN-FPNR method obtains significant lower roughness index value throughout the calibration process, this fact further proves that our proposed FPNR method remove the FPN noise more efficiently.

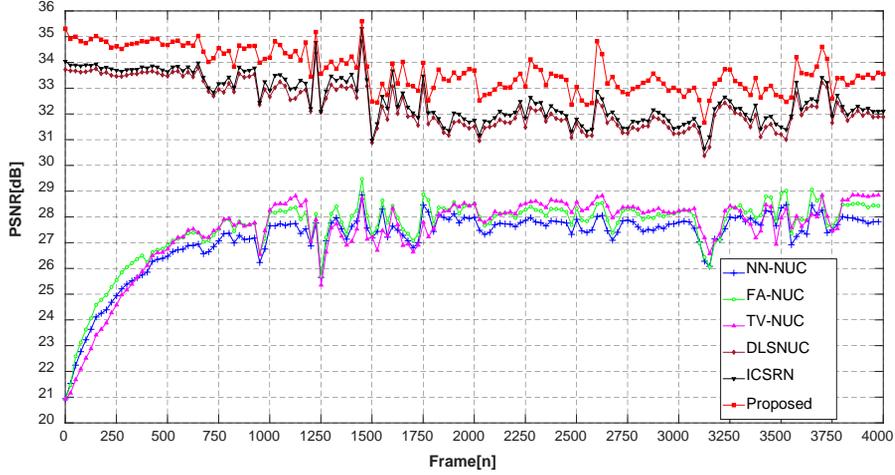

**Figure 8.** PSNR of various FPNR methods for artificially corrupted sequence 1.

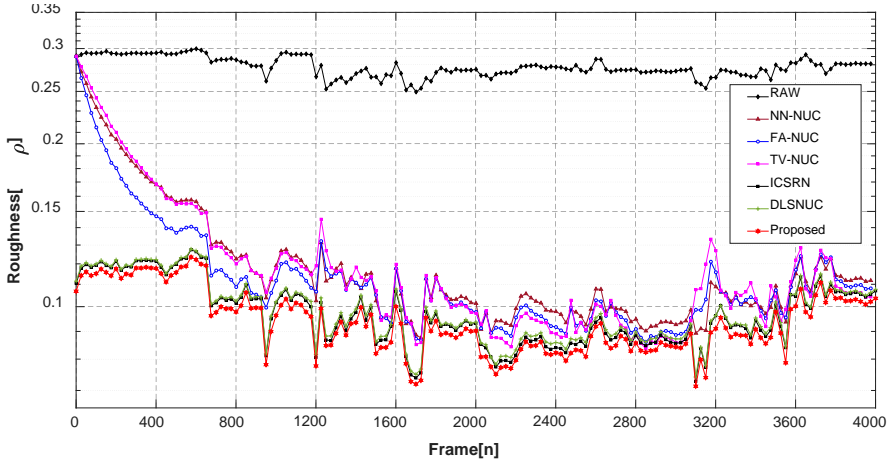

**Figure 9.** Roughness index of various FPNR methods for artificially corrupted sequence 1.

The visual effects of various FPNR methods are shown in Figure 10(c)-(h). From Figure 10(c)-(e), we can observe blurred edges and details as well as more or less residual FPN in the outputs of NN-FPNR, FA-FPNR and TV-FPNR method. Even though the DLSNUC and ICSRN method preserve more image details than traditional SB-FPNR, it is struggle to remove the FPN thoroughly. The strong stripe noise residue area is marked out with red rectangle. In contrast, the proposed CNN-FPNR method can simultaneously suppress the FPN and recover sharp edges and yields visually pleasant results. The main reason is that the deep learning based CNN-FPNR method has the ability to integrally extract and fuse multi-grained features as well as separate the noise and scene detail features precisely from the training process, which helps to predict more accurate calibration parameters to preserve details while smooth out noise.

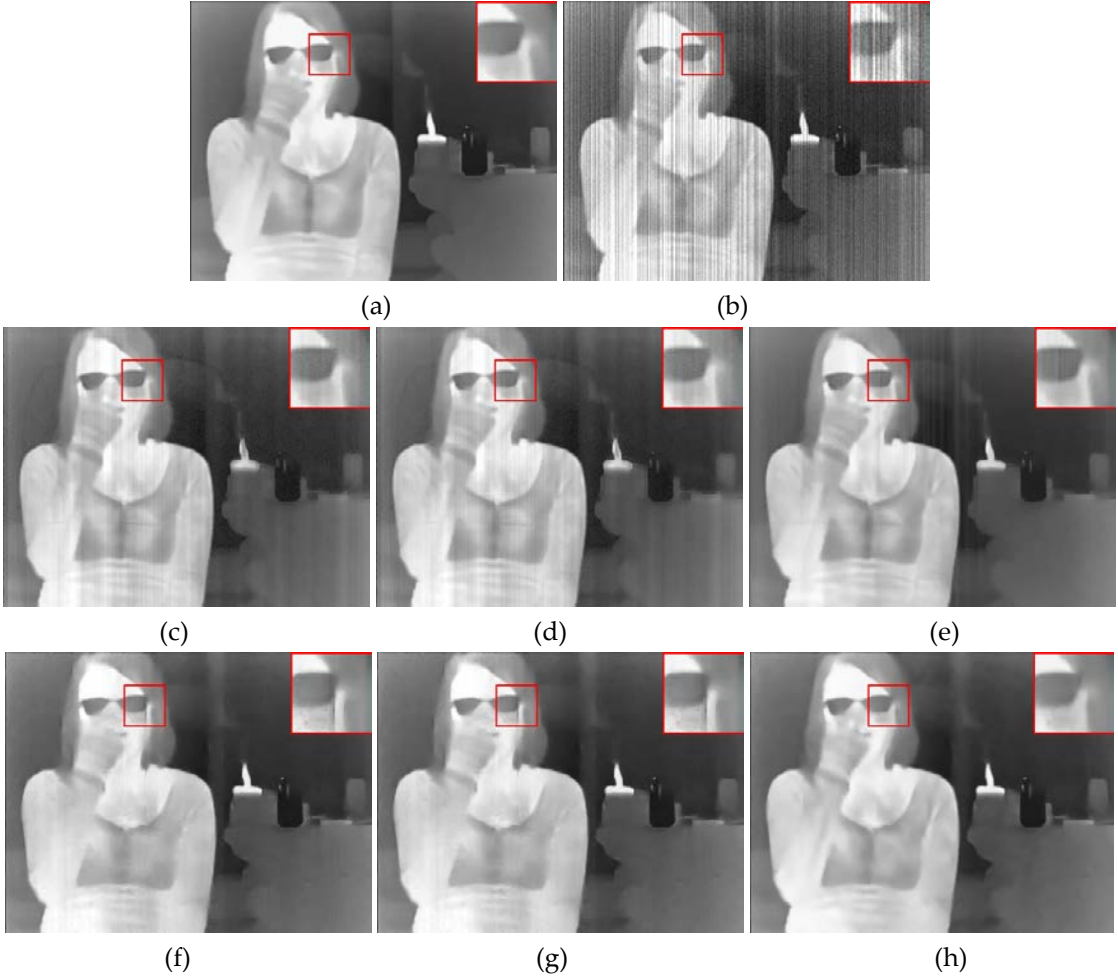

**Figure 10.** Comparison of calibration results for artificially corrupted data in sequence 1. (a) Ground truth 2300th frame; (b) Artificially corrupted 2300th frame; (c) Results of NN-FPNR; (d) Results of FA-FPNR; (e) Results of TV-FPNR; (f) Results of DLSNUC; (g) Results of ICSRN; (h) Results of CNN-FPNR. The red box marks out the region easily to observe stripe noise residue.

In addition, we further implement visual effect comparison on sequence 2, which provides the outdoor building scene involving sharp edges. The results of various FPNR methods are shown in Figure 11. As can be seen, the NN-FPNR, FA-FPNR, and TV-FPNR still suffer from the noise residue artifacts in the corrected results, the DLSNUC and ICSRN produce obvious detail loss on the window frame which is marked out with red rectangle. In contrast, the proposed CNN-FPNR method more thoroughly suppresses the FPN without destroying the intrinsic structure of the image, which yields more sharp and clear details. In conclusion, the proposed CNN-FPNR model exhibits the most prominent performance in various scenarios.

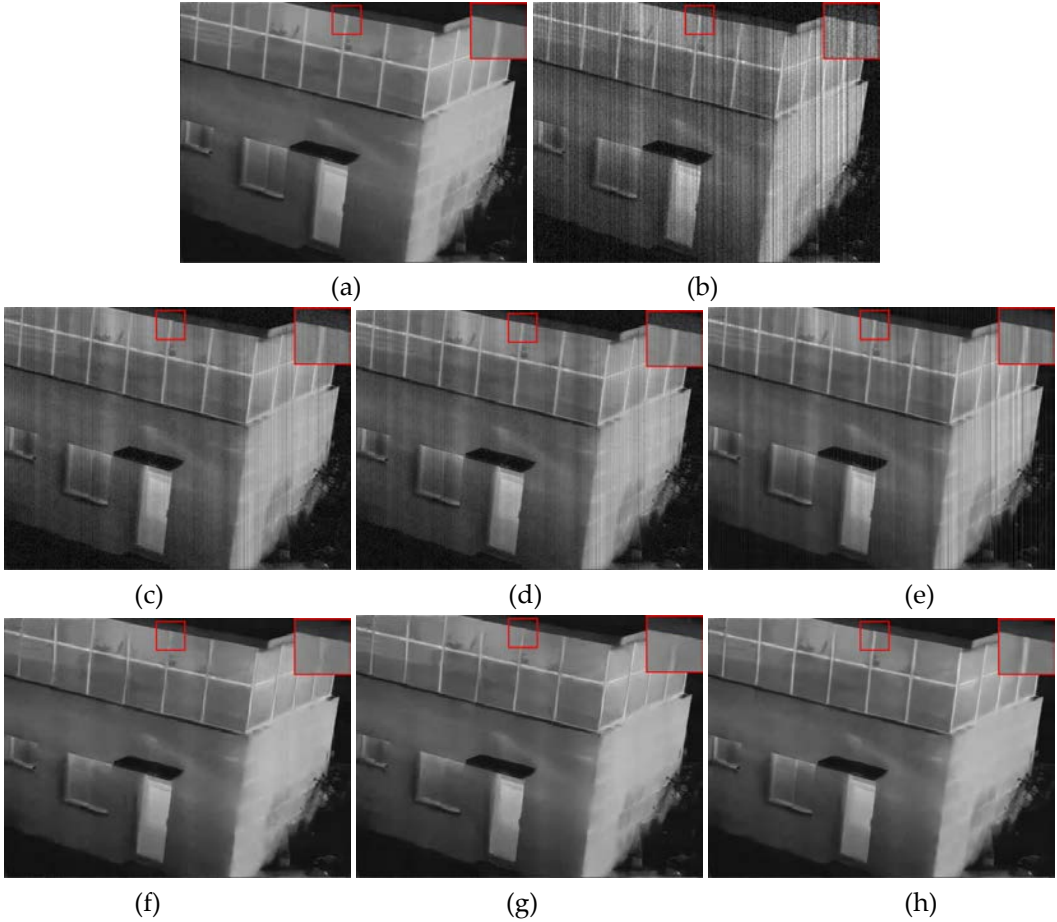

**Figure 11.** Comparison of calibration results for artificially corrupted data in sequence 2. (a) Ground truth 706th frame; (b) Artificially corrupted 706th frame; (c) Results of NN-FPNR; (d) Results of FA-FPNR; (e) Results of TV-FPNR; (f) Results of DLSNUC; (g) Results of ICSRN; (h) Results of CNN-FPNR. The red box marks out the region involving strong edge.

### 4.3.2. Comparison of Deghosting Performance

Traditional SB-FPNR methods are motion-dependent and easily affected by extreme scene, which makes most of them face the problem of ghosting artifacts. These artifacts are generally generated when global scene or partial content of the scene slows down or abruptly halts. In this section, the deghosting performance of different FPNR methods is qualitatively evaluated with artificially corrupted sequence 1. The calibration results of various FPNR methods for 1210th frame are shown in Figure 12(c)-(h), respectively. From Figure 12(c) and (e), we can clearly observe serious ghosting artifacts, which are mainly resulted from the inaccurate estimate of the calibration parameters by using local mean filter. In view of this, the TV-FPNR suppresses most of the ghosting artifacts by minimizing the total variation penalty but still remains obvious residual noise in its output. As shown in Figure 12(f)-(g), the DLSNUC and ICSRN methods overcome the ghosting artifacts but residue some dense stripe and excessively smooth the details (such as the wrinkle marked out with red rectangle). In contrast, our proposed CNN-FPNR method produces more accurate calibration results without any ghosting artifacts and preserves more image details.

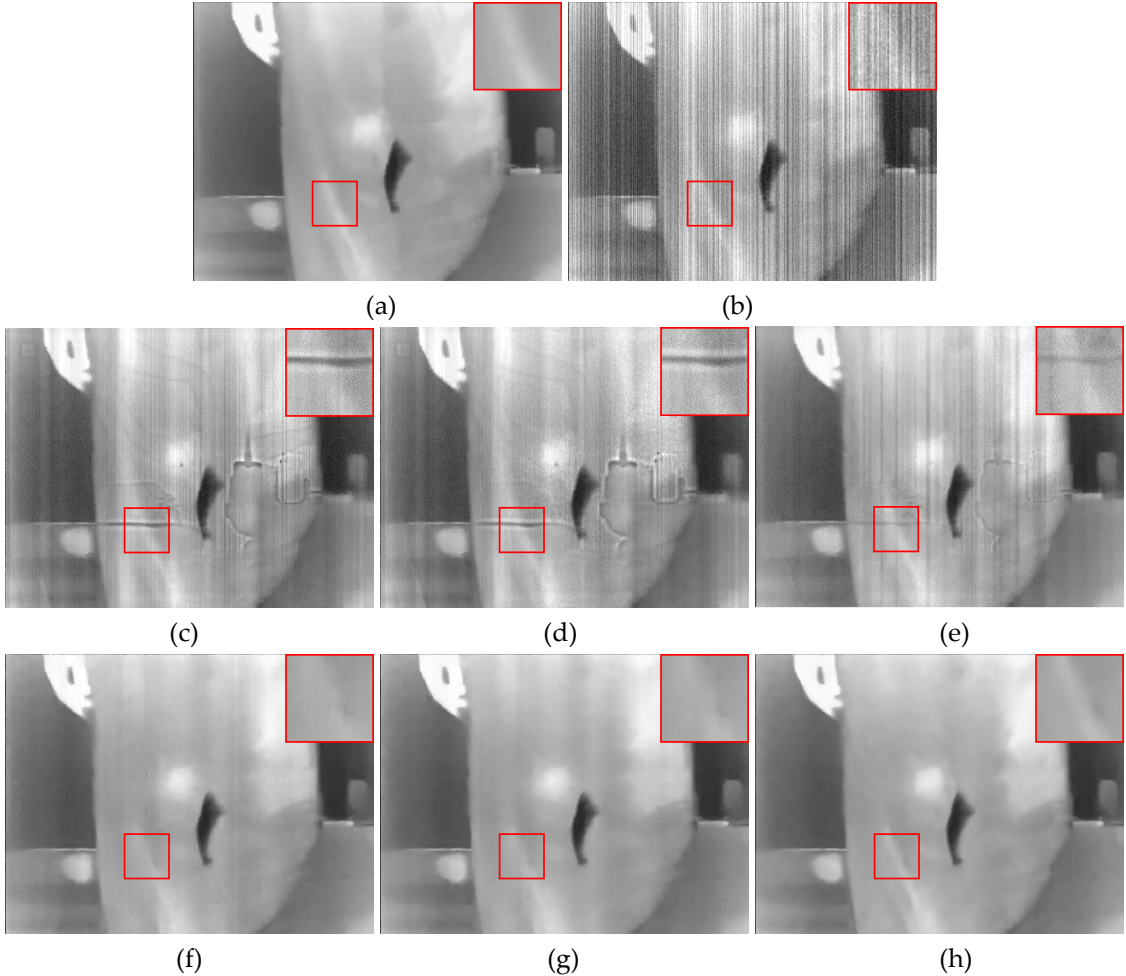

**Figure 12.** Calibration results for varies FPNR methods of corrupted 1210th frame in sequence 1. (a) Ground truth image; (b) Artificially Corrupted image; (c) Result of NN-FPNR; (d) Result of FA-FPNR; (e) Results of TV-FPNR; (f) Results of DLSNUC; (g) Results of ICSRN; (h) Results of CNN-FPNR. The red box marks out the region involving the texture.

## 4.4. Results on Real Infrared Sequences

To further verify the practical effectiveness of different methods, we adopt two real noisy infrared sequences called sequence 3 and sequence 4 to contrast the performance of proposed CNN-FPNR with NN-FPNR, FA-FPNR, TV-FPNR, DLSNUC and ICSRN. Since the ground truth radiation is not available for the observation, we carry out the performance test with the objective roughness index ($\rho$) and subjective visual effect.

### 4.4.1. Visual effect of various FPNR methods

Figure 13(a) illuminates the 270th raw frame in sequence 3, Figure 13(b)-(d) respectively shows the calibration results of NN-FPNR, FA-FPNR, and TV-FPNR method. As shown in Figure 13(b) and (c), the NN-FPNR and FA-FPNR suffer from serious ghosting artifacts in the motion trajectory of object as well as the residual FPN in calibration results. Although the TV-FPNR method eliminates ghosting artifacts to some extent, it produces obvious staircase artifacts in the smooth background region and erases the obvious edge in the area marked out with red rectangle. Moreover, the DLSNUC and ICSRN methods prevent the ghosting artifacts but still blur the edge. In contrast, the proposed CNN-FPNR method suppresses the FPN more thoroughly and preserves more scene details, which yields visually pleasant results without any ghosting and other annoying artifacts.

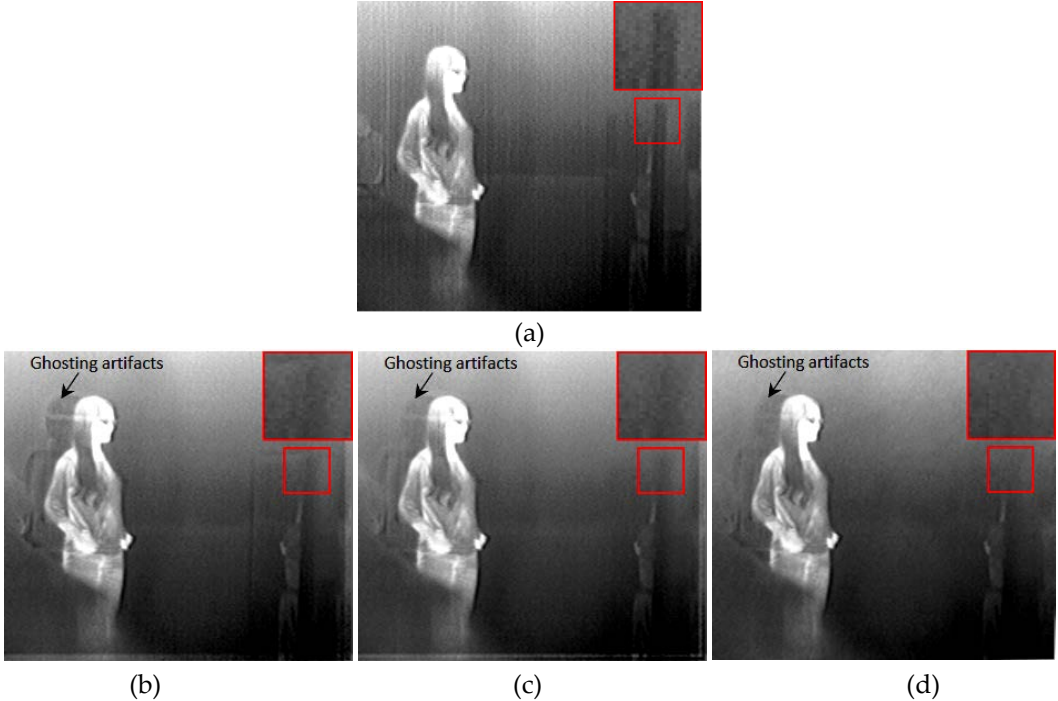

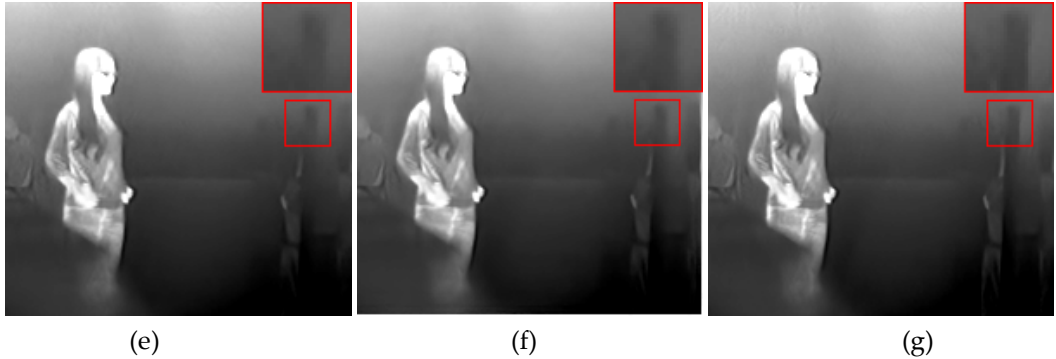

(e)          (f)          (g)

**Figure 13.** Calibrated results of varies FPNR method for the 270th frame in real infrared sequence 3. (a) Raw image; (b) Result of NN-FPNR; (c) Result of FA-FPNR; (d) Result of TV-FPNR; (e) Results of DLSNUC; (f) Results of ICSRN; (g) Result of CNN-FPNR. The red box marks out the region involving obvious edge.

*4.4.2. Comparison of Over Smooth Suppression*

As for traditional SB-FPNR methods, over smooth effects generally arise when the scene has halted for quit a time, but the calibration parameters are still iteratively updating. In order to verify the over smooth suppression capability, we respectively apply NN-FPNR, FA-FPNR, TV-FPNR, DLSNUC, ICSRN and CNN-FPNR upon real noisy sequence 4 and take the calibration results of 1046th frame shown in Figure 14 as an example. As can be seen from Figure 14 (b) - (d), NN-FPNR method produces the most serious over smooth effects and FA-FPNR method preserves most of the obvious edges but still damages the textures severely by adopting the adaptive learning rate strategy. Figure 14 (d) shows that the TV-FPNR method further suppresses the over smooth effect but fails to remove the stripe FPN which marked out in the red box. Although the deep learning based DLSNUC and ICSRN smooth the FPN effectively, it over smooth the image details shown in the red box. In contrast, our proposed CNN-FPNR method simultaneously preserves rich high-frequency details and removes the FPN more thoroughly.

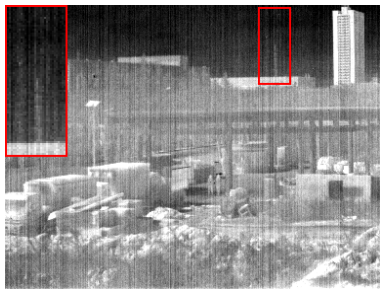

(a)

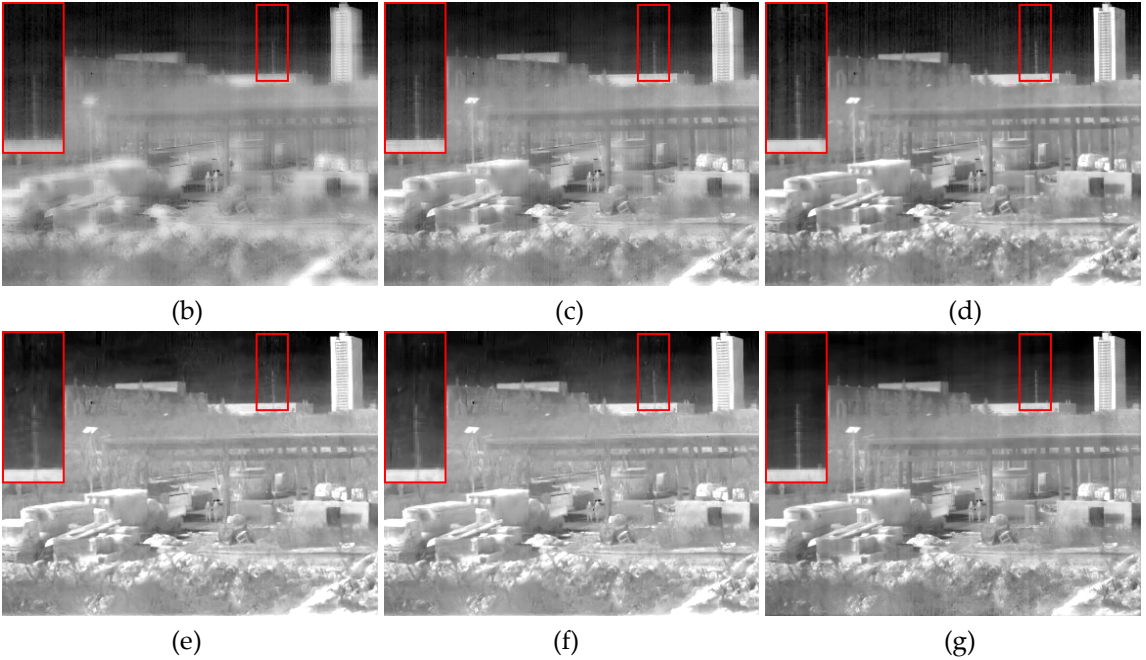

**Figure 14.** Calibrated results of varies FPNR method for 1046th frame in real infrared sequence 4. (a) Raw image; (b) Result of NN-FPNR; (c) Result of FA-FPNR; (d) Result of TV-FPNR; (e) Results of DLSNUC; (f) Results of ICSRN; (g) Result of CNN-FPNR.

*4.4.3. Quantitative assessment of various FPNR methods*

The performance assessment for the averaged values of the whole calibrated frames belonging to each of FPNR methods are listed in Table 4. In this assessment, the SB-FPNR methods were tuned to pursuit the best performance with a trade-off between convergence speed and stability. Furthermore, the gain noise calibration parameter and offset noise calibration parameter of SB-FPNR methods were initialized with one and zero, respectively. From the testing results, we can see clearly that the traditional SB-FPNR methods obtain the relative larger roughness. In addition, the DLSNUC and ICSRN further suppress the FPN, which is affirmed by smaller roughness. In contract, the proposed CNN-FPNR method obtains the smallest reported mean roughness value among all of the FPNR methods. This fact reveals that the proposed CNN-FPNR method suppresses the FPN more effectively in practical application.

**Table 4.** Mean roughness index ($\rho$) for real sequences.

| Sequence | Raw Image | Corrected Image | | | | | |
|---|---|---|---|---|---|---|---|
| | | NN-FPNR | FA-FPNR | TV-FPNR | DLSNUC | ICSRN | Proposed |
| Sequence 3 | 0.2433 | 0.1373 | 0.1355 | 0.1338 | 0.1034 | 0.1019 | **0.1003** |
| Sequence 4 | 0.3083 | 0.1833 | 0.1751 | 0.1621 | 0.1677 | 0.1553 | **0.1513** |

## 5. Conclusion

In summary, an innovative cascade CNN based FPNR method was proposed. Unlike traditional FPNR methods which determine the calibration parameters via reference radiation or inter frame motion, the proposed CNN-FPNR method can directly implement the single frame blind FPNR task without complicated parameter tuning. With special coarse-fine convolution (CF-Conv) unit, the proposed CNN model picks more complete cross-grained features to improve the calibration accuracy. In addition, an elaborate spatial-channel noise attention unit (SCNAU) is presented to further separate FPN and real scene related features, which maximizes the noise reduction and detail preservation effect. The experimental results on both of simulated and real infrared data validate that our proposed CNN-FPNR method yields outstanding precision and sharper visual effect without perceptible ghosting artifacts and over smooth effects.

**Acknowledgements**

This work was supported by Natural Science Foundation of China (NSFC) (Grant Nos. 61674120, 61571338, U1709218, 61672131), Fundamental Research Funds for the Central Universities of China (Grant Nos. JBG161113, 300102328110), and the Key Research and Development Plan of Shaanxi Province (Grant No. 2017ZDCXL-GY-05-01).

**Conflict of interest**

The authors declare that there is no conflict of interest.